\documentclass[twocolumn,pre,aps,showpacs,a4paper,floatfix]{revtex4-1}
\usepackage{graphicx}
\usepackage{psfrag}
\usepackage{subfigure}
\usepackage{epsfig}
\usepackage{calc}
\usepackage{bm}
\usepackage{longtable}
\usepackage{ulem}
\usepackage{color}
\graphicspath{{figures/}}

\newcommand{\beq}{\begin{equation}}
\newcommand{\eeq}{\end{equation}}
\newcommand{\beqa}{\begin{eqnarray}}
\newcommand{\eeqa}{\end{eqnarray}}

\begin{document}

\title{Fluctuation Pressure of Biomembranes in Planar Confinement}
\author{
Thorsten Auth}
\affiliation{
Theoretical Soft Matter and Biophysics,
Institute of Complex Systems and Institute for Advanced Simulation,
Forschungszentrum J\"ulich, 52425 J\"ulich, Germany
}
\author{
Gerhard Gompper}
\affiliation{
Theoretical Soft Matter and Biophysics,
Institute of Complex Systems and Institute for Advanced Simulation,
Forschungszentrum J\"ulich, 52425 J\"ulich, Germany
}


\begin{abstract}
The fluctuation pressure of a lipid-bilayer membrane is important for
the stability of lamellar phases and the adhesion of membranes to surfaces.
In contrast to many theoretical studies, which
predict a decrease of the pressure with the cubed inverse distance
between the membranes, Freund suggested very recently 
a linear inverse distance dependence [Proc.~Natl.~Acad.~Sci.~U.S.A. 
{\bf 110}, 2047 (2013)]. We address this discrepancy by 
performing Monte Carlo simulations for a membrane model discretized on 
a square lattice and employ the wall theorem
to evaluate the pressure for a single membrane between parallel walls.
For distances that are small compared with the lattice constant, the
pressure indeed depends on the inverse distance as predicted by Freund. For
intermediate distances, the pressure depends on the cubed inverse distance
as predicted by Helfrich [Z.~Naturforsch.~A {\bf 33}, 305 (1978)]. Here, 
the crossover length between the two
regimes is a molecular length scale. Finally, for distances large compared
with the mean squared fluctuations of the membrane, the entire membrane
acts as a soft particle and the pressure on the walls again depends linearly
on the inverse distance.
\end{abstract}

\maketitle

\section{Introduction}
The entropic pressure of fluctuating membranes in confinement 
is the main reason for the finite distance between membranes in a swollen
lamellar phase. It plays in important role for membrane adhesion and
many other properties of membranes in confined geometry \cite{lipowsky95}. 
The prediction of the fluctuation pressure of membranes by Helfrich 
\cite{helfrich78} with a $D^{-3}$ dependence on the distance $2D$ between
two confining walls
has been one of the important early successes for the description of 
fluid membranes by curvature elasticity. However, in a very recent 
theoretical and numerical study, Freund \cite{freund13} concluded that
Helfrichs analysis is incorrect, and predicts instead a $D^{-1}$ 
dependence of the fluctuation pressure. In addition, 
Sharma \cite{sharma13} has suggested that this prediction should be taken 
as a stimulus for a new set of experiments.                

The prediction of Freund is surprising, because several  
computer-simulation \cite{janke89,gompper89,netz95a},  
theoretical \cite{kleinert99,bachmann99,kastening06}, and 
experimental studies \cite{safinya86,roux88} seem to have confirmed
Helfrich's prediction very well.
From an application point of view, lamellar phases of lipid bilayers have 
been used, for example, to measure
the effect of additives, such as proteins or polymers, on the interaction
between membranes \cite{bouglet99,brooks93,giahi07}; the interpretation of these
results depends strongly on the distance-dependence of the fluctuation
pressure.
Therefore, a profound understanding of lamellar membrane stacks is very 
important both from a theoretical and an experimental point of view.

\section{Model and Method}

Calculations of the fluctuation pressure are based on the curvature 
elastic energy for nearly planar fluid membranes in the Monge representation,
\begin{equation}
E_b = \frac{\kappa}{2} \int d^2r [ \nabla^2 h({\bf r}) ]^2 \, ,
\label{eq:hamiltonian}
\end{equation}
where $h({\bf r})$ measures the vertical displacement of the membrane
from a planar reference state at horizontal position ${\bf r}$. Here,
$\kappa$ is the bending rigidity on the membrane.
The walls restrict the height variables within the range 
$0 \le h({\bf r}) \le 2D$,
so that $2D$ is the wall separation and $D$ the average distance of
the membrane from the wall.

Helfrich predicted the free energy
change per unit area due to confinement \cite{helfrich78}
\begin{equation}
\label{eq:helf}
\Delta f(D) = c_1 \frac{(k_BT)^2}{\kappa} D^{-2} \, ,
\end{equation}
which implies a fluctuation pressure
\begin{equation}
\label{eq:Helf_pressure}
p= - \left[ \frac{d{\Delta f}}{dD} \right] = 2 c_1 \frac{(k_BT)^2}{\kappa} D^{-3} \, .
\end{equation}
The universal constant $c_1$ was estimated by Helfrich \cite{helfrich78}
to be $c_1=3/32=0.094$.
Subsequent Monte Carlo simulation studies employed a discretized 
membrane model,
in which continuous height variables $h_{i,j}$ are placed on a $N\times N$
square lattice with lattice constant $a$ with periodic 
boundary conditions, so that
\begin{equation}
\label{eq:discretized_bending}
E_b = \frac{\kappa}{2a^2} \sum_{i,j}
[ h_{i+1,j} + h_{i-1,j} +h_{i,j+1} +h_{i,j-1} - 4h_{i,j} ]^2 \, ,
\end{equation}
with the standard discretization of the Laplacian.
These simulations gave results consistent with 
Eq.~(\ref{eq:helf}) and yielded the more precise prediction 
$c_1=0.080$ \cite{gompper89,janke89}.

\begin{figure*}[tp]
  \begin{center}
    \includegraphics[width=1.0\textwidth]{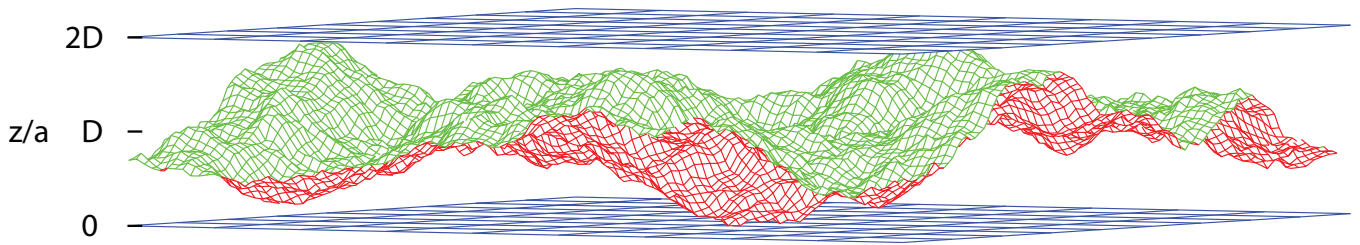}
    \caption{(Color online)
             Membrane with $N=100$ fluctuating between two walls at heights
             $z=0$ and $z=2D=10 a$. The membrane is drawn as a mesh that is
             colored green (light gray) on its upper and red (dark gray) 
             on its lower side. 
             }
    \label{fig:snapshot}
  \end{center}
\end{figure*}

We employ Metropolis Monte Carlo simulations for the model defined by 
Eq.~(\ref{eq:discretized_bending}) with $0 \le h({\bf r}) \le 2D$.
From the simulated membrane conformations, we evaluate the density
distribution $\rho(z)$ of the membrane between two parallel walls.
Here, the density profile is normalized such that
\begin{equation}
\int_0^{2D} dz \, \rho(z) = L^2
\end{equation} 
where $L=Na$ is the linear membrane size.
A typical membrane conformation obtained from a simulation
with $N=100$ is shown in Fig.~\ref{fig:snapshot}. We calculate the 
fluctuation pressure directly from the membrane density profile $\rho(z)$, 
which are related by the wall theorem \cite{hansen2006,gompper91b}, 
\begin{equation}
p=k_BT\rho_w \, ,
\end{equation}
where $\rho_w$ is the density at the wall.

\section{Results}

\begin{figure}[tp]
  \begin{center}
    \includegraphics[width=0.95\columnwidth]{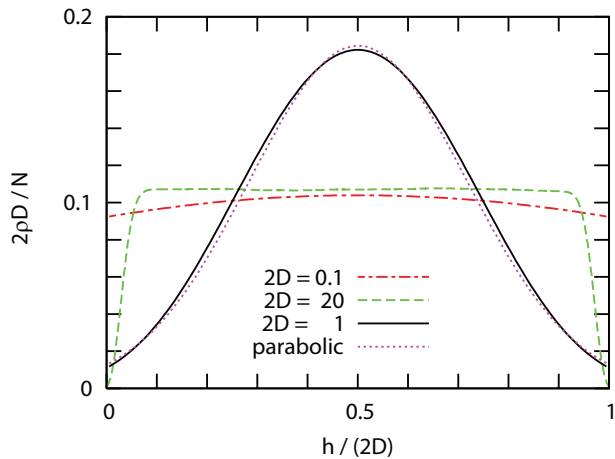}
    \caption{(Color online)
             Membrane density profile $\rho(z)$ between two walls at
             $z=0$ and $z=2D$. The bending rigidity is $\kappa/k_BT=5$.
             The density is normalized by the number of
             lattice sites where the membrane is defined and the separation
             of the walls is given in the legend. The density
             distribution for $2D/a=1$ is compared with that of a membrane
             that fluctuates in a parabolic potential.
             }
    \label{fig:density}
  \end{center}
\end{figure}

In Fig.~\ref{fig:density}, the membrane density profile $\rho(z)$ is shown for
a `small', an `intermediate', an a `large' distance of the parallel walls.
The boundaries between these three regimes will be discussed and quantified
below.
For small distances, the density varies only very slightly as function of
the vertical position $z$ between the walls. 
For intermediate distances, the density profile is 
described very well by a Gaussian function
\cite{leibler87,gompper91a}, see
Fig.~\ref{fig:density}. The membrane that fluctuates between hard walls
can be approximated by a membrane that fluctuates in a parabolic
potential \cite{leibler87}, where an additional term $V(h) = (K/2) h^2$ is added to
Eq.~(\ref{eq:hamiltonian}). The potential strength for the parabolic potential
has been chosen to be $K=(k_B T)^2/(\kappa b^4 D^4)$ where $b \approx 1.14$.
Thus $\langle h^2 \rangle \approx 0.16 D^2$, which is consistent with 
Ref.~\cite{gompper91a}.  For large distances, the 
density profile is flat in the bulk and decreases towards the walls,
as expected for a soft particle. The size of the effective soft particle
is given by the typical mean squared fluctuations of the membrane, 
$\langle h^2 \rangle^{1/2} = (4 \pi^{3/2})^{-1} L (k_BT/\kappa)^{1/2}$, of
the membrane of linear size $L$. 

\begin{figure}[tp]
  \begin{center}
    \includegraphics[width=0.95\columnwidth]{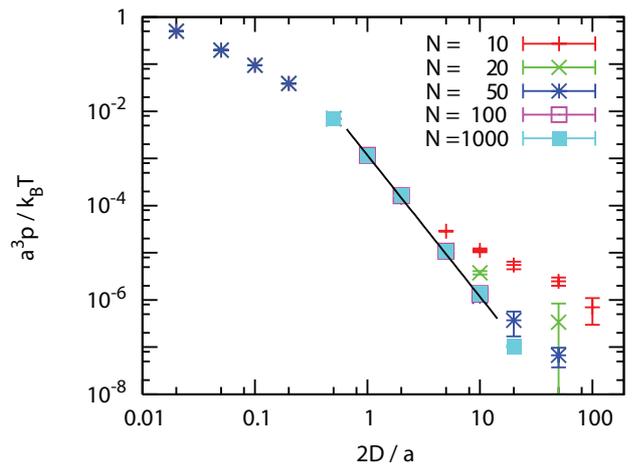}
    \caption{(Color online)
             Membrane pressure $p$ as function of the distance $2D$ between
             the walls for fixed $a$, $\kappa = 5 \, k_B T$, and several
             membrane sizes $N$. The line is a fit to 
             Eq.~(\ref{eq:Helf_pressure}) with the fit parameter $c_1$.
             }
    \label{fig:pressure-L}
  \end{center}
\end{figure}

\begin{figure}[tp]
  \begin{center}
    \includegraphics[width=0.95\columnwidth]{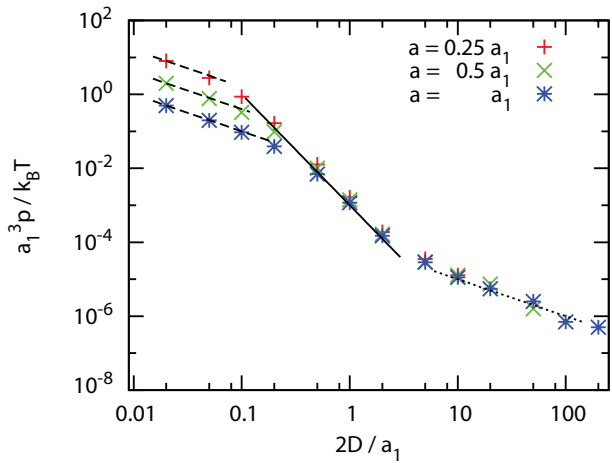}
    \caption{(Color online)
             Membrane pressure $p$ as function of the distance $2D$ between
             the walls. Two systems with smaller discretization length are
             compared with a reference system with $L_1 = 10 a_1$ for
             fixed $L$ and $\kappa = 5 \, k_B T$.  
             The lines indicate the power laws of Eqs.~(\ref{eq:pressure_ideal}), 
             (\ref{eq:Helf_pressure}), and (\ref{eq:finite_size}), by dashed, 
             solid, and dotted lines, respectively. 
             }
    \label{fig:pressure-a}
  \end{center}
\end{figure}

\begin{figure}[tp]
  \begin{center}
    \includegraphics[width=0.95\columnwidth]{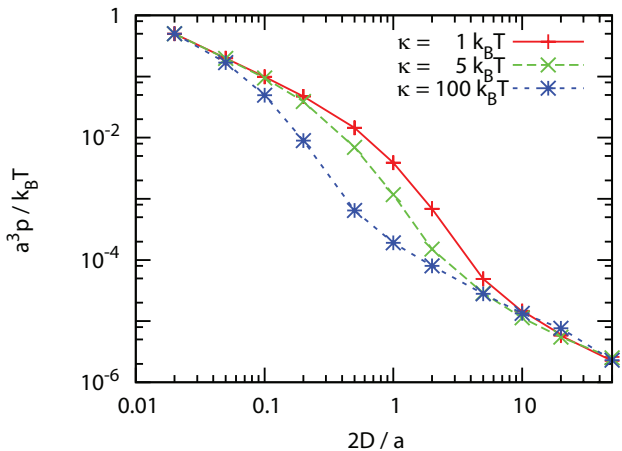}
    \caption{(Color online)
             Membrane pressure $p$ as function of the distance $2D$ between
             the walls for a membrane with $N=10$, fixed $a$ and various
             bending rigidities $\kappa$. Lines are guides to 
             the eye.
             }
    \label{fig:pressure-k}
  \end{center}
\end{figure}

The resulting dependence of the fluctuation pressure on the wall distance
is shown in Figs.~\ref{fig:pressure-L}, \ref{fig:pressure-a}, and
\ref{fig:pressure-k} for various system sizes $N$, discretization
lengths $a$, and bending rigidities $\kappa$. Analogously to the density
distributions in Fig.~\ref{fig:density}, we find three power-law regimes:
a small-$D$ regime with $p\sim D^{-1}$, an intermediate-$D$ regime with
$p\sim D^{-3}$, and a large-$D$ regime with $p\sim D^{-1}$. The fluctuation
pressure for small wall distances does not depend on the membrane
size $N$ and on the bending rigidity $\kappa$. It is given by
\begin{equation}
\label{eq:pressure_ideal}
p=k_BT/(2 D a^2) \, ,
\end{equation}
which is the ideal-gas pressure of $N^2$ independently fluctuating 
height variables, in agreement with Ref.~\cite{freund13}.  
In the regime of intermediate wall distances, the fluctuation pressure
follows the Helfrich prediction (\ref{eq:Helf_pressure}), as can best be
seen in Fig.~\ref{fig:pressure-L}, where the pressure dependence is 
shown for different system sizes $N$. A fit of Eq.~(\ref{eq:Helf_pressure})
gives the universal amplitude $c_1=0.081 \pm 0.002 $, in agreement with
previous simulation results \cite{gompper89,janke89}. 
Note that in this regime the pressure is independent of the discretization 
length $a$. 
For large distances, the pressure is dominated by the translational
degree of freedom of the entire membrane, so that
\begin{equation}
\label{eq:finite_size}
p=\alpha k_BT/(2 D L^2)
\end{equation}
with a function $\alpha(\kappa) = 1+{\cal O}(\sqrt{k_BT/\kappa})$. The
parameter $\alpha$ depends only weakly on the bending rigidity,
see Fig.~\ref{fig:pressure-k}. For further analysis we use $\alpha = 1$.

We can look at the simulation results from three different 
perspectives. For fixed $\kappa$ and $a$, the finite-size regime shifts
to larger $D$ for increased $L$. The finite-size regime is reached when
the lateral correlation length $\xi_\parallel$ reaches the system size $L$. 
Here, $\xi_\parallel$ is defined by decay of the height-height 
correlation function 
$\langle h({\bf r})h({\bf r}')\rangle \sim \exp(-|{\bf r-r}'|/\xi_\parallel)$,
and can be interpreted as the average size of the largest membrane humps,
see Fig.~\ref{fig:snapshot}. It has been shown that 
$\xi_\parallel \sim (\kappa/k_BT)^{1/2} D$ \cite{leibler87,lipowsky95}.
The crossover system size can also be
obtained by equating Eqs.~(\ref{eq:Helf_pressure})
and (\ref{eq:finite_size}), which gives the more precise estimate 
\begin{equation}
D_{\rm fs} = (2 c_1/\alpha)^{1/2} \, L \, (k_BT/\kappa)^{1/2} 
\end{equation} 
with prefactor $(2 c_1/\alpha)^{1/2} \approx 0.4$.
Alternatively, we can consider a system of fixed lateral size $L=Na$
and fixed bending rigidity $\kappa$, and vary the discretization length
$a$ (and accordingly the number of $N^2$ height variables). 
This implies that the prefactor $\kappa/a^2$ in the 
discretized curvature energy in Eq.~(\ref{eq:discretized_bending}) also 
varies.
The results are shown in Fig.~\ref{fig:pressure-a}. In this case,
the finite-size and the Helfrich regime are unaffected by decreasing $a$,
but the small-$D$ regime moves to smaller and smaller values of $D$.
This implies that the small-$D$ regime vanishes in the continuum limit.
The regime of independently fluctuating height variables is entered with
decreasing $D$ when the
lateral correlation length $\xi_\parallel \sim (\kappa/k_BT)^{1/2} D$
of the continuum model drops below the lattice constant $a$, so
that $D_{\rm mol} \sim (k_BT/\kappa)^{1/2} a$. A more precise estimate can
again be obtained by equating Eqs.~(\ref{eq:pressure_ideal}) and 
(\ref{eq:Helf_pressure}), which yields
\begin{equation}
D_{\rm mol} = (2 c_1)^{1/2} \, a \, (k_BT/\kappa)^{1/2} 
\end{equation}
with prefactor $(2 c_1)^{1/2} \approx 0.4$.
This result also shows that the discretization crossover
shifts to smaller values of $D$ with increasing $\kappa$, as shown
explicitly in Fig.~\ref{fig:pressure-k}.

\section{Summary and Conclusions}

We have performed large-scale Monte Carlo simulations
with up to $1,000,000$ lattice sites to clarify the discrepancy between
the predictions of Freund \cite{freund13} and Helfrich \cite{helfrich78}
for the fluctuation pressure of membranes. 
We employ a discretized version of the membrane curvature energy
with short-distance cutoff $a$. This length scale is not an artifact
of the simulation approach, but can be seen as a way to mimic the 
break-down of the continuum description on a molecular level, where
the motion of individual lipid molecules becomes important \cite{goetz99,lindahl00}. 
These
molecular motions have often been denoted as ``protrusion modes". 
We want to emphasize that the discretized curvature model is not a
very good model for the protrusion regime, but should still capture
the crossover between the two regimes.

The main result of our simulations is that although the fluctuation 
pressure of a membrane indeed follows a $D^{-1}$ behavior for very 
small $D$, the curvature elasticity of the membrane 
plays no role in this regime and the pressure is generated by 
independent motion of individual ``molecules".  
The small-$D$ regime occurs when the height differences of neighboring
lattice sites are too small to be controlled by the bending rigidity.
The wall distance $2 D_{\rm mol}$, below which the pressure
shows a $D^{-1}$ behavior, is always smaller than the cutoff length 
$a$ (for $\kappa/k_BT \ge 1$) and decreases with increasing 
$\kappa$.  
For wall distances larger than the molecular length scale, we confirm
the $D^{-3}$ distance dependance of the pressure predicted by Helfrich,
with the universal amplitude $c_1=0.080$. 
This curvature-elasticity controlled behavior is seen in 
the regime $a < \xi_\parallel < L$.

Interestingly, when the data in Fig.~4 of Ref.~\cite{freund13} is 
replotted in a double-logarithmic presentation, a crossover becomes
visible from $p\sim D^{-1}$ to a faster decay. 
The value of the crossover
distance should depend on the number of fluctuation modes employed in
the analysis. We find that the length scale $\lambda$ in the model
proposed by Freund corresponds to our molecular cutoff length $a$. In contrast,
Sharma \cite{sharma13} claims that $\lambda$ is large compared with the
molecular size and therefore the use of the continuum model is
justified.

In conclusion, the analysis of previous experiments on the basis
of the theoretical expression for the fluctuation pressure with
an inverse cubic dependence on the confinement remains
valid --- for wall distances corresponding to parallel correlation 
lengths
in the range $a < \xi_\parallel < L$. For smaller $D$ (with
$\xi_\parallel < a$), we confirm an inverse linear dependence of the
fluctuation pressure. 
Therefore, it would be very interesting to perform new
experiments and molecular simulations to investigate the
crossover from the undulation- to the protrusion-dominated regime.

\acknowledgments
Stimulating discussions with S. Dietrich (Stuttgart) are gratefully
acknowledged.


\begin{thebibliography}{23}
\expandafter\ifx\csname natexlab\endcsname\relax\def\natexlab#1{#1}\fi
\expandafter\ifx\csname bibnamefont\endcsname\relax
  \def\bibnamefont#1{#1}\fi
\expandafter\ifx\csname bibfnamefont\endcsname\relax
  \def\bibfnamefont#1{#1}\fi
\expandafter\ifx\csname citenamefont\endcsname\relax
  \def\citenamefont#1{#1}\fi
\expandafter\ifx\csname url\endcsname\relax
  \def\url#1{\texttt{#1}}\fi
\expandafter\ifx\csname urlprefix\endcsname\relax\def\urlprefix{URL }\fi
\providecommand{\bibinfo}[2]{#2}
\providecommand{\eprint}[2][]{\url{#2}}


\bibitem[{\citenamefont{{Lipowsky} and {Sackmann}}(1995)}]{lipowsky95}
\bibinfo{editor}{\bibfnamefont{R.}~\bibnamefont{{Lipowsky}}} \bibnamefont{and}
  \bibinfo{editor}{\bibfnamefont{E.}~\bibnamefont{{Sackmann}}}, eds.,
  {\bibinfo{title}{Structure and dynamics of membranes - from cells to
  vesicles}}, vol.~\bibinfo{volume}{1} of {\bibinfo{series}{Handbook of
  Biological Physics}} (\bibinfo{publisher}{Elsevier},
  \bibinfo{address}{Amsterdam}, \bibinfo{year}{1995}).

\bibitem[{\citenamefont{Helfrich}(1978)}]{helfrich78}
\bibinfo{author}{\bibfnamefont{W.}~\bibnamefont{Helfrich}},
  \bibinfo{journal}{Z. Naturforsch. A} \textbf{\bibinfo{volume}{33}},
  \bibinfo{pages}{305} (\bibinfo{year}{1978}).

\bibitem[{\citenamefont{Freund}(2013)}]{freund13}
\bibinfo{author}{\bibfnamefont{L.}~\bibnamefont{Freund}},
  \bibinfo{journal}{Proc. Natl. Acad. Sci. U.S.A.}
  \textbf{\bibinfo{volume}{110}}, \bibinfo{pages}{2047} (\bibinfo{year}{2013}).

\bibitem[{\citenamefont{Sharma}(2013)}]{sharma13}
\bibinfo{author}{\bibfnamefont{P.}~\bibnamefont{Sharma}},
  \bibinfo{journal}{Proc. Natl. Acad. Sci. U.S.A.}
  \textbf{\bibinfo{volume}{110}}, \bibinfo{pages}{1976} (\bibinfo{year}{2013}).

\bibitem[{\citenamefont{Janke et~al.}(1989)\citenamefont{Janke, Kleinert, and
  Meinhart}}]{janke89}
\bibinfo{author}{\bibfnamefont{W.}~\bibnamefont{Janke}},
  \bibinfo{author}{\bibfnamefont{H.}~\bibnamefont{Kleinert}}, \bibnamefont{and}
  \bibinfo{author}{\bibfnamefont{M.}~\bibnamefont{Meinhart}},
  \bibinfo{journal}{Phys. Lett. B} \textbf{\bibinfo{volume}{217}},
  \bibinfo{pages}{525} (\bibinfo{year}{1989}).

\bibitem[{\citenamefont{Gompper and Kroll}(1989)}]{gompper89}
\bibinfo{author}{\bibfnamefont{G.}~\bibnamefont{Gompper}} \bibnamefont{and}
  \bibinfo{author}{\bibfnamefont{D.}~\bibnamefont{Kroll}},
  \bibinfo{journal}{Europhys. Lett.} \textbf{\bibinfo{volume}{9}},
  \bibinfo{pages}{59} (\bibinfo{year}{1989}).

\bibitem[{\citenamefont{Netz and Lipowsky}(1995)}]{netz95a}
\bibinfo{author}{\bibfnamefont{R.~R.} \bibnamefont{Netz}} \bibnamefont{and}
  \bibinfo{author}{\bibfnamefont{R.}~\bibnamefont{Lipowsky}},
  \bibinfo{journal}{Europhys. Lett.} \textbf{\bibinfo{volume}{29}},
  \bibinfo{pages}{345} (\bibinfo{year}{1995}).

\bibitem[{\citenamefont{Kleinert}(1999)}]{kleinert99}
\bibinfo{author}{\bibfnamefont{H.}~\bibnamefont{Kleinert}},
  \bibinfo{journal}{Phys. Lett. A} \textbf{\bibinfo{volume}{257}},
  \bibinfo{pages}{269} (\bibinfo{year}{1999}).

\bibitem[{\citenamefont{Bachmann et~al.}(1999)\citenamefont{Bachmann, Kleinert,
  and Pelster}}]{bachmann99}
\bibinfo{author}{\bibfnamefont{M.}~\bibnamefont{Bachmann}},
  \bibinfo{author}{\bibfnamefont{H.}~\bibnamefont{Kleinert}}, \bibnamefont{and}
  \bibinfo{author}{\bibfnamefont{A.}~\bibnamefont{Pelster}},
  \bibinfo{journal}{Phys. Lett. A} \textbf{\bibinfo{volume}{261}},
  \bibinfo{pages}{127} (\bibinfo{year}{1999}).

\bibitem[{\citenamefont{Kastening}(2006)}]{kastening06}
\bibinfo{author}{\bibfnamefont{B.}~\bibnamefont{Kastening}},
  \bibinfo{journal}{Phys. Rev. E} \textbf{\bibinfo{volume}{73}},
  \bibinfo{pages}{11101} (\bibinfo{year}{2006}).

\bibitem[{\citenamefont{Safinya et~al.}(1986)\citenamefont{Safinya, Roux,
  Smith, Sinha, Dimon, Clark, and Bellocq}}]{safinya86}
\bibinfo{author}{\bibfnamefont{C.}~\bibnamefont{Safinya}},
  \bibinfo{author}{\bibfnamefont{D.}~\bibnamefont{Roux}},
  \bibinfo{author}{\bibfnamefont{G.}~\bibnamefont{Smith}},
  \bibinfo{author}{\bibfnamefont{S.}~\bibnamefont{Sinha}},
  \bibinfo{author}{\bibfnamefont{P.}~\bibnamefont{Dimon}},
  \bibinfo{author}{\bibfnamefont{N.}~\bibnamefont{Clark}}, \bibnamefont{and}
  \bibinfo{author}{\bibfnamefont{A.}~\bibnamefont{Bellocq}},
  \bibinfo{journal}{Phys. Rev. Lett.} \textbf{\bibinfo{volume}{57}},
  \bibinfo{pages}{2718} (\bibinfo{year}{1986}).

\bibitem[{\citenamefont{Roux and Safinya}(1988)}]{roux88}
\bibinfo{author}{\bibfnamefont{D.}~\bibnamefont{Roux}} \bibnamefont{and}
  \bibinfo{author}{\bibfnamefont{C.}~\bibnamefont{Safinya}},
  \bibinfo{journal}{J. Phys. (France)} \textbf{\bibinfo{volume}{49}},
  \bibinfo{pages}{307} (\bibinfo{year}{1988}).

\bibitem[{\citenamefont{Bouglet and Ligoure}(1999)}]{bouglet99}
\bibinfo{author}{\bibfnamefont{G.}~\bibnamefont{Bouglet}} \bibnamefont{and}
  \bibinfo{author}{\bibfnamefont{C.}~\bibnamefont{Ligoure}},
  \bibinfo{journal}{Eur. Phys. J. B} \textbf{\bibinfo{volume}{9}},
  \bibinfo{pages}{137} (\bibinfo{year}{1999}).

\bibitem[{\citenamefont{Brooks and Cates}(1993)}]{brooks93}
\bibinfo{author}{\bibfnamefont{J.}~\bibnamefont{Brooks}} \bibnamefont{and}
  \bibinfo{author}{\bibfnamefont{M.}~\bibnamefont{Cates}}, \bibinfo{journal}{J.
  Chem. Phys.} \textbf{\bibinfo{volume}{99}}, \bibinfo{pages}{5467}
  (\bibinfo{year}{1993}).

\bibitem[{\citenamefont{Giahi et~al.}(2007)\citenamefont{Giahi, Faris,
  Bassereau, and Salditt}}]{giahi07}
\bibinfo{author}{\bibfnamefont{A.}~\bibnamefont{Giahi}},
  \bibinfo{author}{\bibfnamefont{M.~E.~A.} \bibnamefont{Faris}},
  \bibinfo{author}{\bibfnamefont{P.}~\bibnamefont{Bassereau}},
  \bibnamefont{and} \bibinfo{author}{\bibfnamefont{T.}~\bibnamefont{Salditt}},
  \bibinfo{journal}{Eur. Phys. J. E} \textbf{\bibinfo{volume}{23}},
  \bibinfo{pages}{431} (\bibinfo{year}{2007}).

\bibitem[{\citenamefont{{Hansen} and {McDonald}}(2006)}]{hansen2006}
\bibinfo{author}{\bibfnamefont{J.~P.} \bibnamefont{{Hansen}}} \bibnamefont{and}
  \bibinfo{author}{\bibfnamefont{I.~R.} \bibnamefont{{McDonald}}},
  {\bibinfo{title}{Theory of Simple Liquids}}
  (\bibinfo{publisher}{Academic Press}, \bibinfo{address}{London},
  \bibinfo{year}{2006}), \bibinfo{edition}{3rd} ed.

\bibitem[{\citenamefont{Gompper and Kroll}(1991{\natexlab{a}})}]{gompper91b}
\bibinfo{author}{\bibfnamefont{G.}~\bibnamefont{Gompper}} \bibnamefont{and}
  \bibinfo{author}{\bibfnamefont{D.~M.} \bibnamefont{Kroll}},
  \bibinfo{journal}{J. Phys. I France} \textbf{\bibinfo{volume}{1}},
  \bibinfo{pages}{1411} (\bibinfo{year}{1991}{\natexlab{a}}).

\bibitem[{\citenamefont{Leibler and Lipowsky}(1987)}]{leibler87}
\bibinfo{author}{\bibfnamefont{S.}~\bibnamefont{Leibler}} \bibnamefont{and}
  \bibinfo{author}{\bibfnamefont{R.}~\bibnamefont{Lipowsky}},
  \bibinfo{journal}{Phys. Rev. B} \textbf{\bibinfo{volume}{35}},
  \bibinfo{pages}{7004} (\bibinfo{year}{1987}).

\bibitem[{\citenamefont{Gompper and Kroll}(1991{\natexlab{b}})}]{gompper91a}
\bibinfo{author}{\bibfnamefont{G.}~\bibnamefont{Gompper}} \bibnamefont{and}
  \bibinfo{author}{\bibfnamefont{D.~M.} \bibnamefont{Kroll}},
  \bibinfo{journal}{Europhys. Lett.} \textbf{\bibinfo{volume}{15}},
  \bibinfo{pages}{783} (\bibinfo{year}{1991}{\natexlab{b}}).

\bibitem[{\citenamefont{Goetz et~al.}(1999)\citenamefont{Goetz, Gompper, and
  Lipowsky}}]{goetz99}
\bibinfo{author}{\bibfnamefont{R.}~\bibnamefont{Goetz}},
  \bibinfo{author}{\bibfnamefont{G.}~\bibnamefont{Gompper}}, \bibnamefont{and}
  \bibinfo{author}{\bibfnamefont{R.}~\bibnamefont{Lipowsky}},
  \bibinfo{journal}{Phys. Rev. Lett.} \textbf{\bibinfo{volume}{82}},
  \bibinfo{pages}{221} (\bibinfo{year}{1999}).

\bibitem[{\citenamefont{Lindahl and Edholm}(2000)}]{lindahl00}
\bibinfo{author}{\bibfnamefont{E.}~\bibnamefont{Lindahl}} \bibnamefont{and}
  \bibinfo{author}{\bibfnamefont{O.}~\bibnamefont{Edholm}},
  \bibinfo{journal}{Biophys. J.} \textbf{\bibinfo{volume}{79}},
  \bibinfo{pages}{426} (\bibinfo{year}{2000}).

\end{thebibliography}

\end{document}